# Deep Convolutional Neural Network to Detect J-UNIWARD

Guanshuo Xu
independent researcher
228 Stewart Ave.
Kearny, NJ 07032
guanshuo.xu@gmail.com

## ABSTRACT
This paper presents an empirical study on applying convolutional neural networks (CNNs) to detecting J-UNIWARD — one of the most secure JPEG steganographic method. Experiments guiding the architectural design of the CNNs have been conducted on the JPEG compressed BOSSBase containing 10,000 covers of size 512×512. Results have verified that both the pooling method and the *depth* of the CNNs are critical for performance. Results have also proved that a 20-layer CNN, in general, outperforms the most sophisticated feature-based methods, but its advantage gradually diminishes on hard-to-detect cases. To show that the performance generalizes to large-scale databases and to different cover sizes, one experiment has been conducted on the CLS-LOC dataset of ImageNet containing more than one million covers cropped to unified size of 256×256. The proposed 20-layer CNN has cut the error achieved by a CNN recently proposed for large-scale JPEG steganalysis by 35%. Source code is available via GitHub: https://github.com/GuanshuoXu/deep_cnn_jpeg_steganalysis

## 1. INTRODUCTION
Current published works on applying convolutional neural networks (CNNs) to image steganalysis mainly focus on detecting steganography embedding in the original spatial domain [1–6]. Studies on applying CNNs for JPEG steganalysis have not been extensively carried out even though JPEG steganography could be more conveniently used in practice. Recently, Zeng *et al.* [7] for large-scale JPEG steganalysis designed a *hybrid* CNN optimized upon various quantized DCT subbands of decompressed input images. It is worth noting that only three convolutional layers exist in their proposed CNN, and average pooling which frequently appear in spatial domain steganalysis [2,4] has also been employed. Experiments have been performed on subsets of the whole ImageNet Database with more than 14 million images. Results demonstrate that their CNN outperform traditional feature-based methods in detecting JPEG steganography provided that the number of training data is huge. However, as will be shown in this study, both the shallow architecture and the average pooling layers are too conservative to fully bring the strength of deep learning into play.

Spatial domain Steganography, as its name indicates, makes changes directly to the pixel values in the original spatial domain. Because of the variation in the locations and values of the changes made on image pixels during information embedding, the CNNs, extremely powerful in mining local patterns, run the risk of memorizing the embedding patterns which would eventually harm the generalization of the trained models. Hence, CNNs are forced to be shallow (5–6 layers [2,4]) for spatial domain steganalysis, and, average pooling, rarely used in computer vision, has also found its value because averaging neighboring elements reduced the risk of memorizing exact embedding locations. In contrast, JPEG steganography makes embedding changes to the quantized DCT coefficients. When transformed back to the spatial domain, the changes made on the DCT coefficients spread to all the pixels in their 8x8 blocks, exposing JPEG steganography more to the fire of deep CNNs compared with its counterpart in the spatial domain.

This paper presents the latest results using CNN to detect the most secure JPEG steganography method — the JPEG version of the universal distortion function (J-UNIWARD) [15]. It has been discovered that the designed all-convolutional 20-layer CNN, equipped with batch normalization and shortcut connections for efficient gradient back-propagation, has generally better performance compared with the most sophisticated feature-based methods, when tested on the BOSSBase compressed with JPEG quality factors of 75 and 95, and embedded by J-UNIWARD using Gibbs simulator with rates of 0.1, 0.2, 0.3, and 0.4 bpnzAC. To show that the performance of the designed CNN generalizes to large-scale databases and to different image sizes, one experiment has been conducted on the CLS-LOC dataset of ImageNet, which contains more than one million covers cropped to 256×256, and recompressed with QF75, then embedded with rate of 0.4 bpnzAC. The proposed 20-layer CNN has cut the error achieved by a CNN recently proposed for large-scale JPEG steganalysis [7] by 35%.

The architecture of the 20-layer CNN will be introduced in Section 2. All the experiments will be presented in Section 3. Section 4 summarizes this paper.

## 2. THE PROPOSED CNN ACHITECTURE
The entire architecture of our proposed CNN is included in Fig. 1. Only the forward pass appears in the figure, as should be enough for understanding the ideas.

Similar to what has been done in the traditional feature-based steganalysis methods [8,9,11], the input in the JPEG format are first transformed to the spatial domain (without the last rounding step), then go through a set of filter banks. In the proposed CNN, undecimated DCT of size 4×4 are selected to project every single input to 16 different frequency bands. The DCT kernels are fixed and not optimized during training. During the initial stage of this study, DCT sizes of 2×2, 3×3, 4×4, 5×5, and 8×8 have been tested; the best results are obtained with size 4×4. Also have been tested are removing the DC subband or the highest frequency subbands, but no obvious improvement have been observed. Other types of filter banks, e.g., the Garbor filters [9], have not been studied yet and should belong to future research. Same as in [7–9,11], in this work, only the magnitudes of the subbands are used, and they are further truncated with a slightly tuned global threshold value of 8. Quantization as another essential ingredient in the traditional feature extraction procedure has been abandoned to prevent unnecessary information loss; after all, the CNNs do not explicitly assemble histograms for statistical modeling. Zeng *et al.* [7] propose to learn a sub-CNN on each of the quantized versions of the subbands, it can be argued that such a design could potentially over-complicate the problem, because the CNN should be able to learn something similar to quantization with better information preserving and gradient-decent friendly operations such as linear scaling and non-linear activations. Nevertheless, truncation to limit the range of input data seems still necessary; it has been observed that without truncation the CNNs experienced slow convergence.

Following these pre-processing steps is the core part of the CNN comprising 20 convolutional layers and a global average pooling layer. This part of the CNN is responsible for learning optimized function to transform each of the pre-processed input into a 384-D feature vector for classification. All the convolutional layers are followed by Batch-Normalization (BN) to reduce *internal covariant shift* [17] and the most widely used Rectified Linear Unit (ReLU) [21] as the non-linear activation function. The convolution kernels have a unified size of 3×3 along spatial dimensions. The width of the CNN is mainly constrained by the GPU memory; we managed to fit this CNN in a single GPU with 12GB memory. It is essential for the CNNs to reduce the spatial resolutions by pooling while going deeper, though attention should be paid that operations with strides skips modes in the 8x8 JPEG blocks, which would cause more information loss in JPEG steganalysis. Let the CNNs take into considerations of the non-stationarity of the input, similar to the mode-wise statistical modeling done in traditional feature-based steganalysis [8,9], could be a valuable future work. In our CNN, pooling is achieved by convolutional layers with stride 2, after which the spatial sizes of data are cut by half and the number of channels doubles. Empirical study carried out in Section 3.1 compares convolution with average and max pooling and suggests a clear advantage for convolution with stride over average and max pooling for JPEG steganalysis; after all, convolutional layers at least introduce more learnable parameters and therefore increase the depth of CNN. Our *deep* CNN contrasts the *shallow* CNN [7] containing only three convolutional layers. The importance of depth will be further verified in Section 3.1.

A 20-layer CNN, even though already equipped with BN and ReLU, still more or less suffers from the *gradient vanishing* problem causing inefficient training, as will be shown in Section 3.1. To overcome this issue, the structure of shortcut connections inspired by [16–20] is brought into play. All the shortcut connections added in the CNN are concisely illustrated in Fig. 1 (Left) as curved arrows. The solid curved arrows denote direct shortcut connections allowing the convolutional layers in the middle to learn only residuals [19,20]; the dashed curved arrows denote transformed shortcut connections elaborated in Figure 1 (Right) because the element-wise addition requires input data of exactly same sizes. With shortcut connections, the depth following the shortest path is only 5, whereas the longest path has 20 layers, achieving both the strength of modeling and efficient training.

The linear classification module following the global pooling layer is simply composed of a fully-connected layer (no more hidden layers and no non-linear activations) and a softmax layer to transform the feature vectors to posterior probabilities for each class. Final class labels are determined by choosing the class corresponding to the larger posteriors.

# 3. EXPERIMENTS

The primary database used in this study is the BOSSbase v1.01 [12] containing 10,000 uncompressed images, initially taken by seven cameras in the RAW format, and transformed to 8-bit grayscale images, then cropped to obtain the size of 512×512. To generate covers for JPEG steganography, the images were compressed with QF75 and QF95 as representatives for low and high quality using Matlab's *imwrite* function. The corresponding stegos were generated through data embedding into the compressed images. Hence, for each classification problem, the dataset contains 10,000 cover–stegos pairs. J-UNIWARD served as the only steganographic method in this study. According to the steganalysis results presented in [8,9,11], J-UNIWARD should be the most secure algorithm embedding in the JPEG domain. Embedding rates of 0.1, 0.2, 0.3, and 0.4 bpnzAC were selected for experiments.

All of the experiments using the CNN reported in this study were performed on a modified version of the Caffe toolbox [14]. Mini-batch stochastic gradient descent was used to solve all the CNNs in experiments. The momentum was fixed to 0.9. The learning rate was initialized to 0.001, and scheduled to decrease 10% every 5000 training iterations. Parameters in the convolution kernels were randomly initialized from zero-mean Gaussian distribution with standard deviation of 0.01; bias learnings were disabled in convolutional layers and fulfilled in the BN layers. Parameters in the last fully-connected layers were initialized using Xavier initialization. Weight decay was only enabled in the final fully-connected layer. A mini-batch of 32 images comprising 16 cover–stego pairs was the input for each training iteration. Each input pair during training were randomly horizontally mirrored and rotated by a multiple of 90 degrees in a synchronized manner for cover-stego to guarantee that the CNN always learns the difference caused by data embedding. The training set was randomly shuffled for each epoch of training. After every multiple of 5000 iterations, the parameters in the CNN were saved.

## 3.1 Deeper is Better

Results in Section 3.1 are obtained with narrower CNNs (2/3 of the width in Fig. 1), on the BOSSBase with embedding rate of 0.4bpnzAC, and QF75 and QF95. We used half of the data for training and the other half for validation. We ran all the experiments 150,000 iterations to ensure the CNNs have enough time to converge.

The first experiment aim to choose the best pooling method from three common options: convolution, average pooling, and max pooling, all with stride 2. Architectures of the CNNs used for comparisons are illustrated in Fig. 2. Please refer to the description under Fig. 1 if there is any confusion with the elements in the figure. Results shown in Fig. 3 demonstrate that pooling with convolution has significantly lower validation errors. Note that pooling with convolution makes the CNN deeper (11 layers versus 6 layers).

Inspired from this, it would be natural to add more layers to further enhance the performance. In this paper we add up to 20 layers. Not surprisingly, adding more layers causing trouble in training; this can be observed in Fig. 4 showing abnormally higher training errors achieved by a more complex 20-layer CNN, and there is no question that the validation performance also suffer. Fortunately, adding shortcut connections (same as displayed in Fig. 1) solves the problem. This can be clearly observed in Fig. 5.

## 3.2 Results on the BOSSbase

Table 1 and Table 2 show the final ensemble results on the BOSSBase using the CNN in Fig. 1. We stopped the training of CNNs after 90000 iterations (288 epochs). Final testing results for each CNN were obtained by averaging the probability output of the test data from the CNNs models saved on their 80000, 85000 and 90000 training iterations; we ensembled four CNNs for each train-test split. The performance of the proposed CNN shown in the first rows of Table 1 and Table 2 clearly outperforms the best feature-based method without even using the information of the selection-channel, but we can also observe that in harder scenarios, e.g., at the rate of 0.2 bpnzAC with QF95, the CNN has almost no advantage. In extreme cases, i.e., 0.1 bpnzAC with QF95, after running for 40,000 iterations the CNNs failed to reduce the training error and we just stopped and put the number 0.5 which means random guess. Instead of training from scratch for all the embedding rates, we also provide our finetuning results by initializing the CNNs with parameters optimized by the tasks of higher embedding rates, which is similar to what have been done in [3]. Better results are observed for QF95, but in the cases of QF75, funetuning failed to improve the performance.

**Table 1: Classification Errors with QF75**

| | Embedding rates (bpnzAC) | | | |
|---|---|---|---|---|
| | **0.1** | **0.2** | **0.3** | **0.4** |
| **Proposed (Fig1)** | 0.3283 | 0.1947 | 0.1124 | 0.0641 |
| **Proposed (finetune)** | 0.3469 | 0.2086 | 0.1141 | 0.0641 |
| **SCA-GFR [11]** | 0.3598 | 0.2316 | 0.1409 | 0.0807 |

**Table 2. Classification Errors with QF95**

| | Embedding rates (bpnzAC) | | | |
|---|---|---|---|---|
| | **0.1** | **0.2** | **0.3** | **0.4** |
| **Proposed (Fig1)** | 0.5000 | 0.3974 | 0.3106 | 0.2364 |
| **Proposed (finetune)** | 0.4554 | 0.3852 | 0.3067 | 0.2364 |
| **SCA-GFR [11]** | 0.4629 | 0.3998 | 0.3303 | 0.2620 |

## 3.3 Results on The CLS-LOC dataset

The CLS-LOC dataset after being cropped to 256×256, and recompressed with QF75 has 1,152,197 covers in the training set, 48,627 covers in the validation set, and 97,296 covers in the testing set. We only tested rate of 0.4 bpnzAC due to constraints on computation facilities. Results are given in Fig. 6, the testing errors obtained by the saved models with best validation results are 0.256 for the CNN in [7] and 0.168 for the proposed CNN. This is a significant performance improvement as the proposed has cut the error by about 35%.

# 4. CONCLUSION

In this paper, a 20-layer CNN has been proposed and tested on both the BOSSBase with cover size of 512×512 and the CLS-LOC dataset with processed cover size of 256×256. It has been demonstrated that deep CNN can beat feature-based methods except in very difficult cases, and is therefore a promising research direction for further performance improvement.

Future works to further move this research ahead includes the following:

1. Replace 4×4 DCT with more effective filter banks or something equivalent.
2. Bring the information caused by pooling (subsampling) back by making the CNN phase-ware.
3. Making the CNN even deeper …
4. Test the proposed CNN on other JPEG steganographic algorithms.

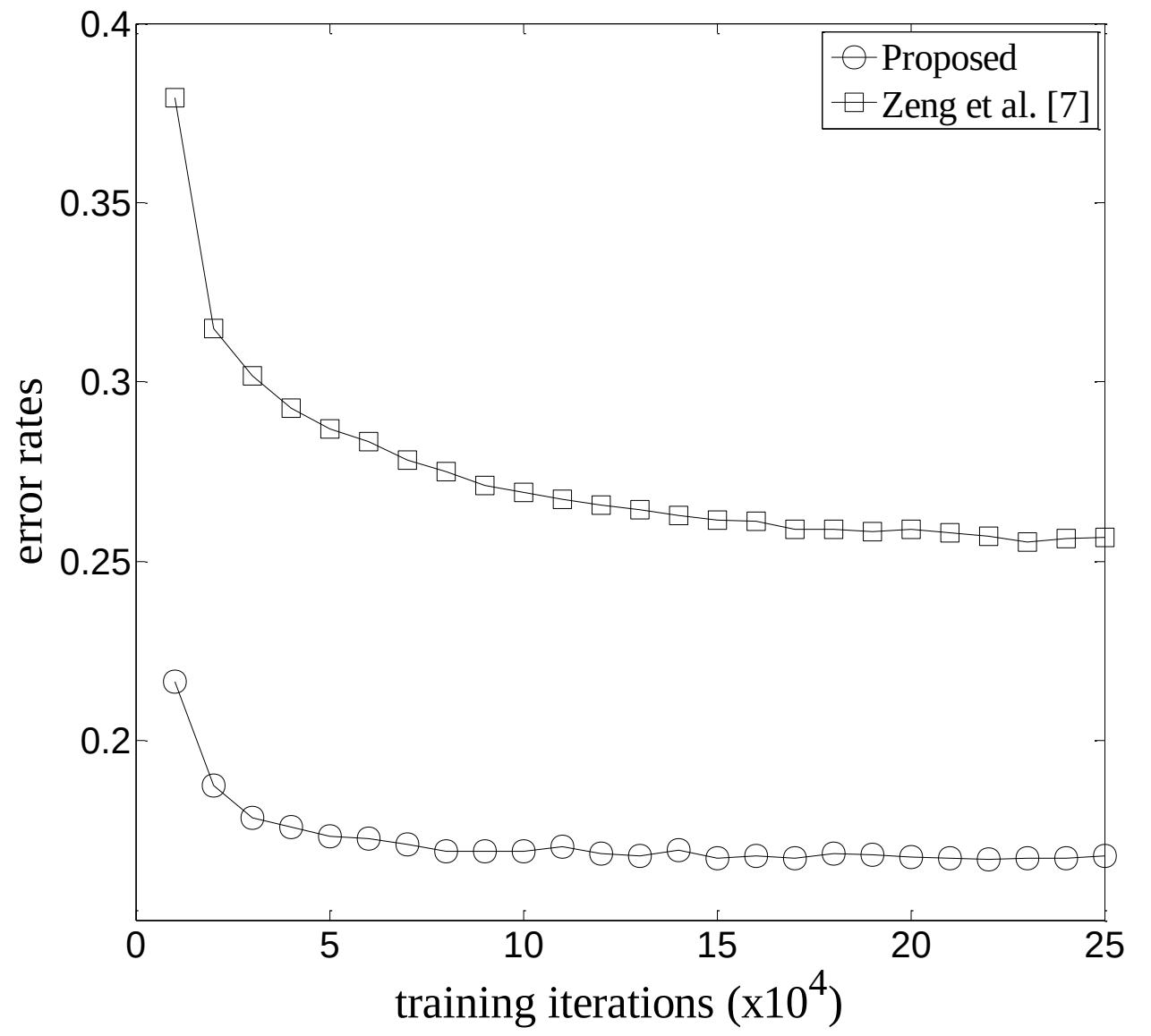

**Figure 6: Comparison of validation errors versus training iterations between the proposed CNN and the CNN in [7] at 0.4bpnzAC embedding rate.**

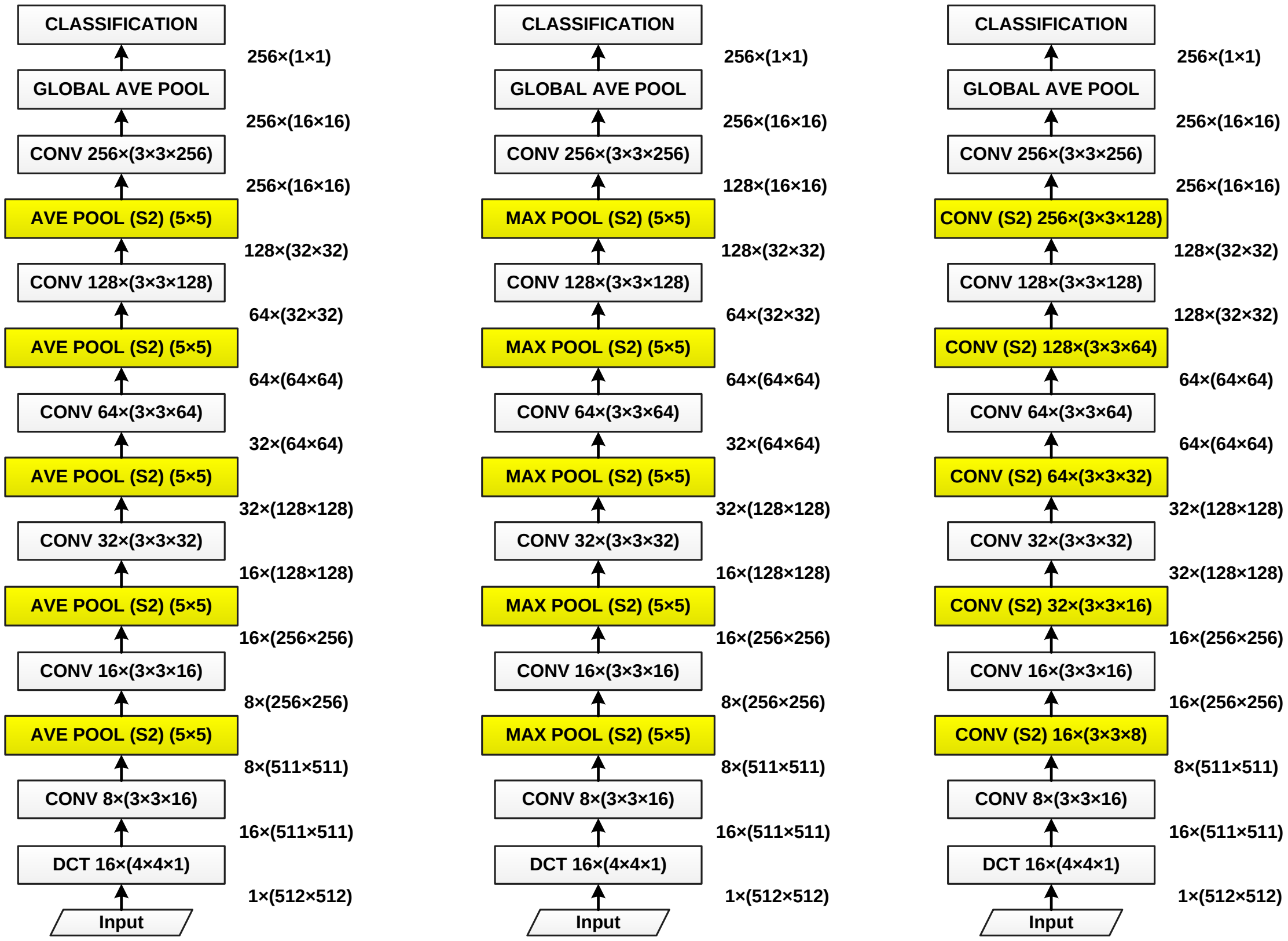

**Figure 2: (Left) A 6-layer CNN equipped with average pooling. (Center) A 6-layer CNN equipped with max pooling. (Right) An 11-layer all convolutional CNN.**

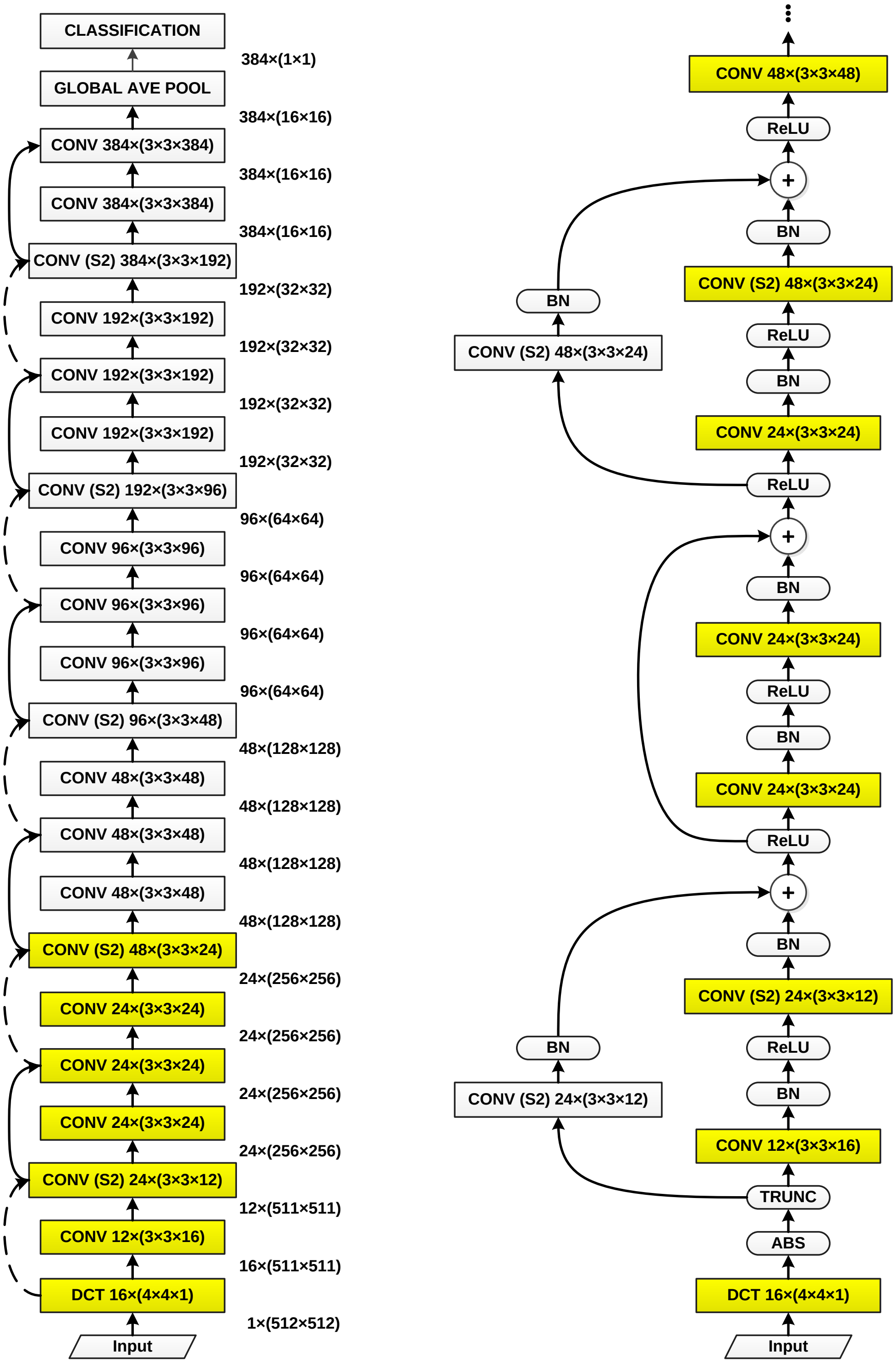

**Figure 1: (Left) The proposed CNN architecture in the concise form. Data sizes following (number of channels) × (height × width) are displayed on the right side. CONV denotes convolutional layers, with kernel sizes following (number of kernels) × (height × width × number of channels). Spatial subsampling is fulfilled by convolution with stride equals 2 (S2). Solid curved arrows denote direct shortcut connections; dashed curved arrows denote transformed shortcut connections. Padding is applied wherever is necessary. (Right) A complete elaboration of the marked portion in the left figure. The *plus* signs indicate element-wise additions.**

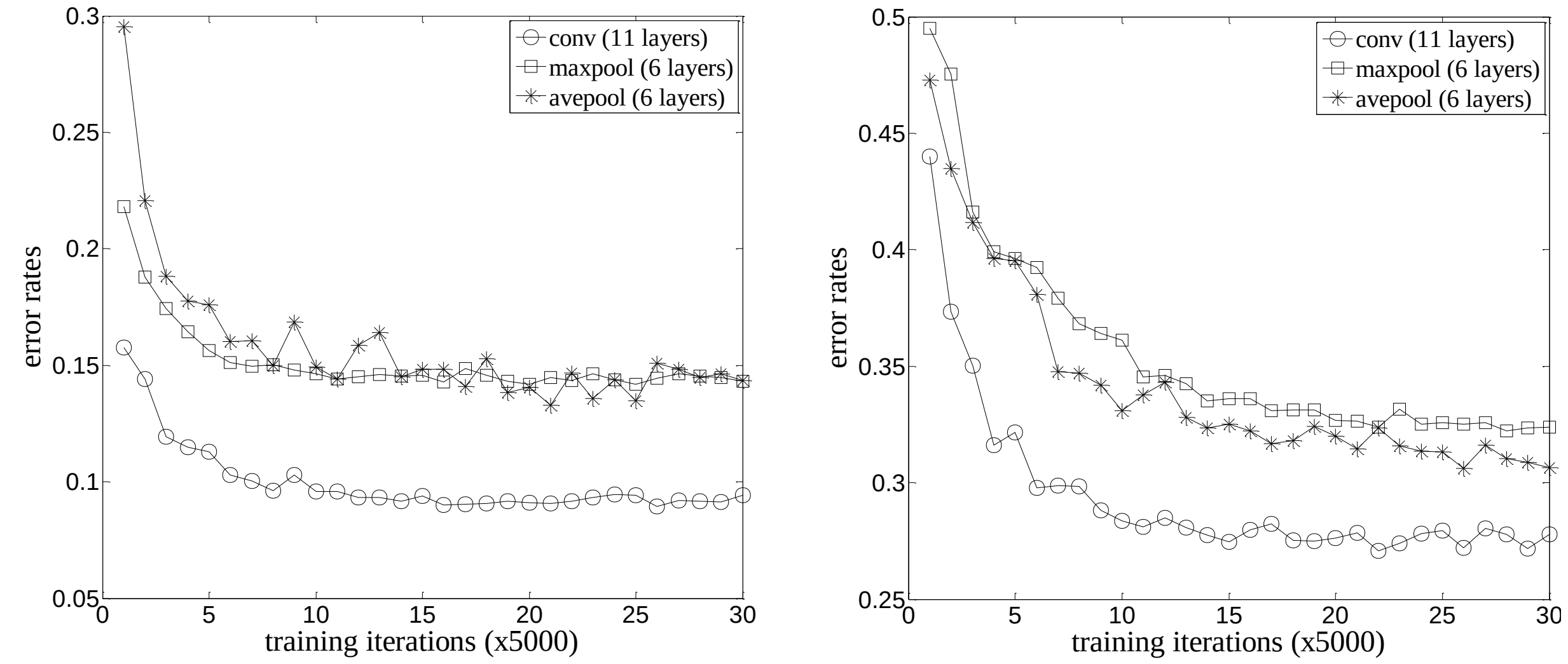

**Figure 3: Comparison of validation errors versus training iterations between three types of pooling (convolution, max, and average) at 0.4bpnzAC embedding rate for (Left) QF75 and (Right) QF95.**

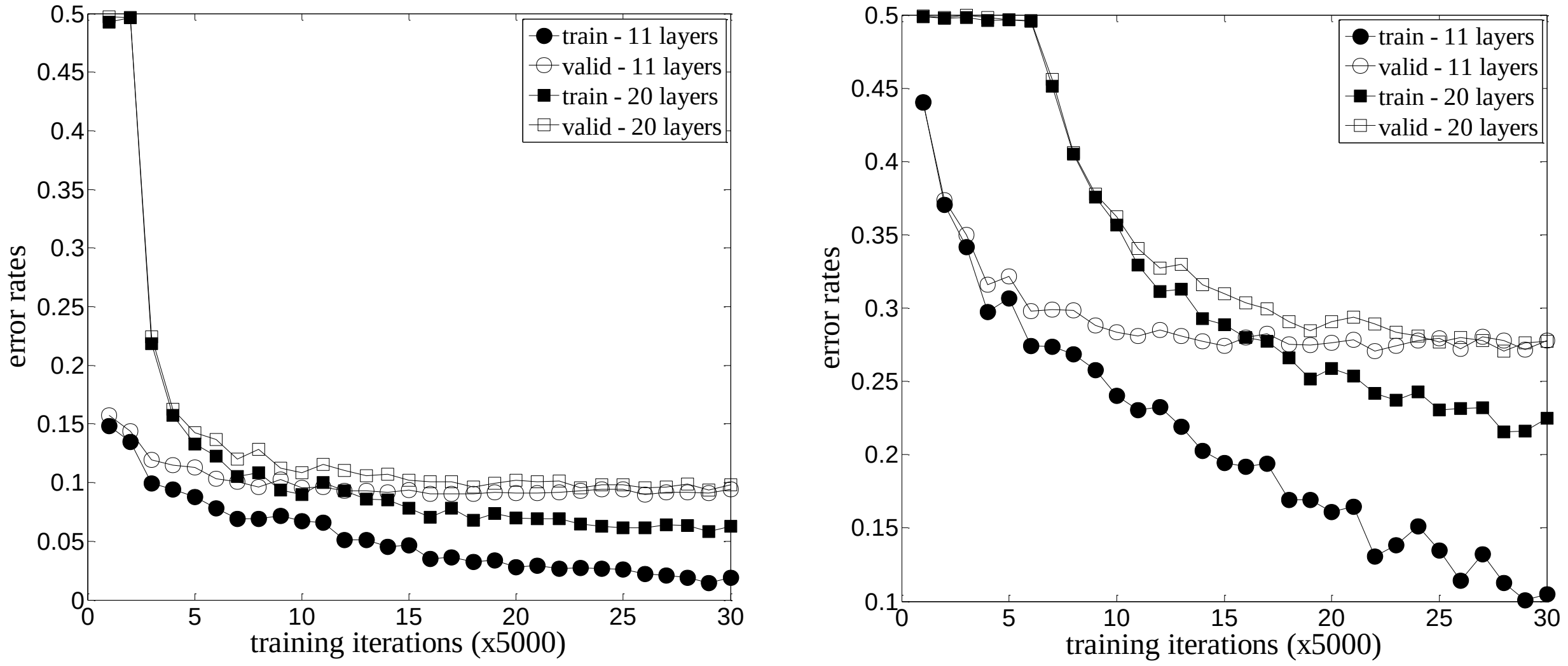

**Figure 4: Comparison of training and validation errors versus training iterations between a 11-layer CNN and a 20-layer CNN without shortcut connections at 0.4bpnzAC embedding rate for (Left) QF75 and (Right) QF95.**

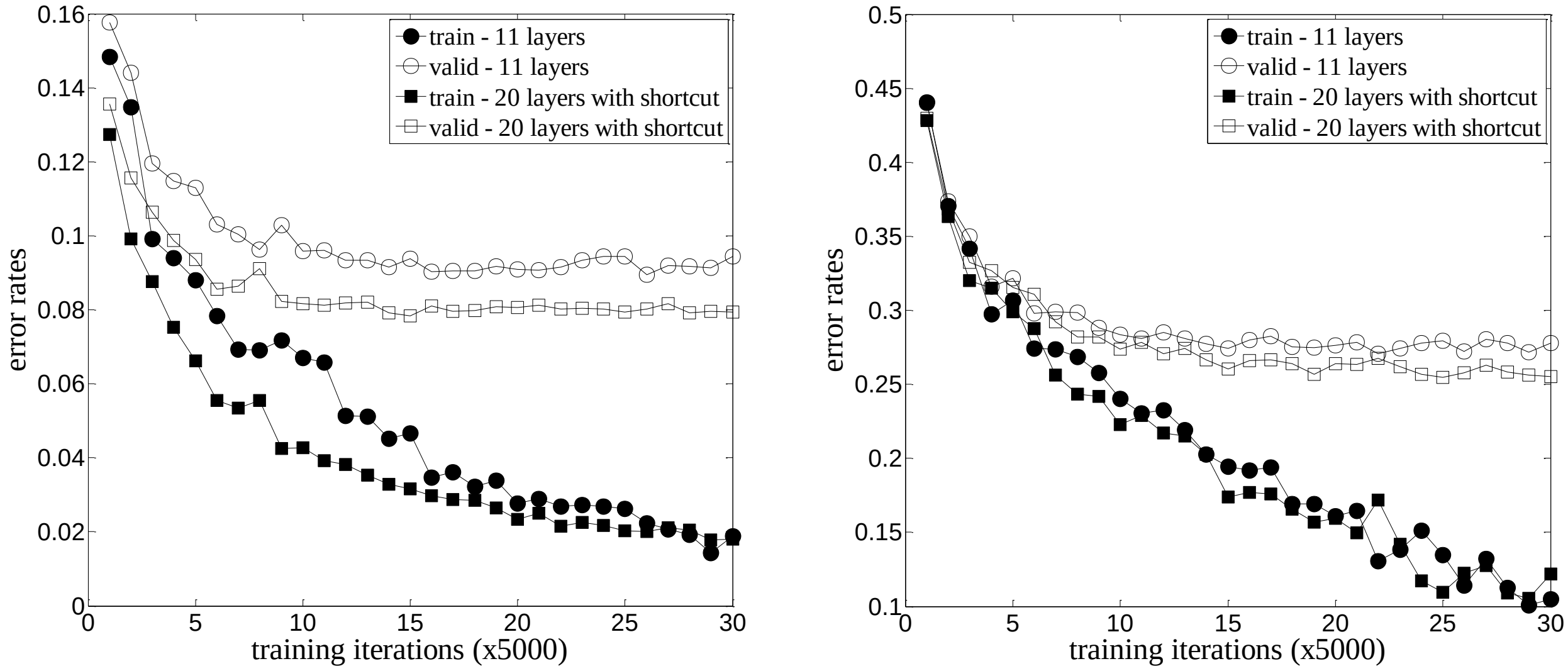

**Figure 5: Comparison of training and validation errors versus training iterations between a 11-layer CNN and a 20-layer CNN with shortcut connections at 0.4bpnzAC embedding rate for (Left) QF75 and (Right) QF95.**